# Impact of Sensing Range on Real-Time Adaptive Control of Signalized Intersections Using Vehicle Trajectory Information

Andalib Shams
Iowa State University
2711 S Loop Dr, Ames, IA 50010, USA
ashams@iastate.edu

Christopher M. Day*
Iowa State University
813 Bissell Rd, Ames, IA 50011, USA
(515) 294-2140 cmday@iastate.edu

*Corresponding Author.

Word Count: 6376 words

**ABSTRACT**

Advanced signal control algorithms are anticipated with the increasing availability of vehicle speed and position data from vehicle-to-infrastructure communication and from sensors. This study examines the impact of the sensing range, meaning the distance from the intersection that such data can be obtained, on the quality of the signal control. Two advanced signal control methods, a Self-Organizing Algorithm (SOA) and Phase Allocation Algorithm (PAA), were implemented in simulation and tested to understand the impact of sensing ranges. SOA is based on fully-actuated control with an added secondary extension for vehicle platoons along the arterial. PAA uses dynamic programming to optimize phase sequences and phase duration within a planning horizon. Three different traffic scenarios were developed: symmetric, asymmetric, and balanced. In general, both algorithms exhibited improvements in performance as the sensing range was increased. Under the symmetric volume scenario, SOA converged at 1000 ft and PAA converged at 1500 ft (with vehicles traveling at 45mph). Under the asymmetrical and balanced volume scenarios, both algorithms outperformed conventional methods. Both algorithms performed better than coordinated-actuated control if the sensing range is 660 ft or higher. For low sensing ranges, SOA experiences similar delay compared to fully-actuated control with advance detector and PAA experienced more delay than coordinated-actuated control in symmetric and asymmetric scenario but performed better for asymmetric scenario. The results suggest that for the SOA algorithm, the sensing range may constrain the maximum allowable secondary extension, hence as the sensing range increases, the vehicular delay would decrease for arterial movements and increase for non-arterial movements. For PAA, arterial and non-arterial delay decreases with increase in sensing range until it converges.

**Keywords:** Traffic control algorithm, vehicle trajectory, advanced infrastructure sensing

## INTRODUCTION

Recent technological innovations have begun to make it increasingly feasible to track the positions and speeds of vehicles at signalized intersections, enabling improved control decisions that will yield safer and more efficient operation. Connected Vehicle (CV) technology is one such innovation. An issue with CV technology is that it is unclear when a sufficient proportion of vehicles will be connected such that enhanced control strategies are enabled. In parallel to this, advances in sensor and detector technology are making it increasingly feasible to passively track vehicles at intersections. This includes systems such as radar, LiDAR, video, and infrared detection. This could potentially create a mechanism equivalent to one-directional communication from vehicles to infrastructure. To name this concept, "Advanced infrastructure sensing" (AIS) may be able to detect near 100% of the vehicle fleet with existing technology while true CV penetration rate may require a long time to reach such levels. Early results of field tests have shown that radar detection shows promise in (*1*)

Regardless of whether CV or AIS technology are used to establish vehicle trajectories, the *sensing range* of the technology will be a critical determinant of the ability of the technology to impact the signal operation. Most existing research in the area of CV-enabled signal control innovations have focused on penetration rate, and the issue of sensing range has not been explored in depth. Conventional vehicle detection consists of point detectors located at a position that facilitates dilemma zone protection, approximately 5 seconds of travel time from the stop bar. It is not known whether a similar distance is sufficient for enhanced signal control applications, or whether a greater distance is needed. There is a need to better understand the impact of the sensing range on control applications.

This study explores the performance envelope of AIS-enabled signal control using microsimulation. The outcomes of advanced signal control algorithms are compared under different sensing ranges. The outcomes of this study will help to identify the scale of potential benefits should this technology be implemented in the real world, and help define the desirable spatial range that should be sought for intersection control applications.

## LITERATURE REVIEW

Research on intelligent signal control goes back to the 1960s. Over time, numerous methods for real-time adaptive control have been described in the literature, but there are two concepts representing alternative approaches to the problem that many of these methods can be categorized under. One is the use of a planning horizon, in which it is supposed that the inbound vehicles expected at an intersection can be arranged into an arrival table, and this information can be used to determine an optimal sequence and duration of green times over a time period on the order of the next few minutes. Another approach is to employ rule-based mechanisms to extend or truncate greens or influence next-phase decisions. In that case, the decisions tend to be simpler but occur on a shorter time frame, similar to actuated control. This analysis has adopted one algorithm from each of these two methods to investigate the impact of sensing range on the signal operation.

### Trajectory-Based Control under a Planning Horizon

Datesh et al. (*2*) proposed a k-means clustering algorithm for a fully decentralized intersection that considers service of queued vehicles and approaching vehicles as separate stages. Assuming 100% market penetration of connected vehicles, this method reduces the delay by 6% over the actuated-coordinated signal control.

Researchers at University of Arizona presented (*3*) a unified platoon-based formulation, named PAMSCOD, to perform online network-level traffic signal optimization for multiple modes of traffic using vehicle-to-infrastructure communication. The method is based on the earlier RHODES traffic control system (*4*, *5*). PAMSCOD determines phase timing every 30 seconds for four cycles in the future. Vehicles send a phase request to the controller along with mode, speed, and position data. Multiple signals are coordinated using virtual priority requests. A 40% penetration rate is critical for the algorithm to operate effectively.

Goodall et al. (*6*) proposed decentralized traffic control logic based on connected vehicle data, using a 15-second horizon using vehicle location, speed, and headway information. At least 25% market penetration was found to be needed to achieve improvement.

Continuing from earlier work at University of Arizona, Feng et al. (*7*) presented a real-time adaptive phase allocation algorithm using CV location and speed data that used bi-level optimization specific to eight-phase control in which the timing of boundaries between "barrier groups" (representing main street or side street) were determined on the upper level and the timing and order of individual phases (i.e., through and left-turn movements) were handled at the lower level. The total delay was reduced significantly for a high penetration rate. A subsequent study (*8*) implemented a system for signal coordination using fixed priority requests. This research was further extended (*9*) to include a method for optimizing signal offsets. This system used a cycle length to maintain coordination but achieved flexible phase timing based on vehicle demand measured through CV methods.

Wang et al. (*10*) used mobile computing data to collect vehicle space-time trajectories and proposed a multi-intersection phase to represent safe vehicle movements across connected intersections. All the intersections are viewed as one integrated intersection within which vehicles can move according to their planned path. Mixed integer linear programming was used to optimize vehicle trajectories and traffic control. A Lagrangian decomposition approach was used to make the optimization problem scalable. The results of this study suggest that it is not essential to maintain a conventional cycle-split-offset approach to maintain coordination, and flexible phase sequence and phase duration can also be used to improve coordination.

Islam and Hajbabaie (*11*) developed a decentralized, hierarchical approach that used an objective of maximizing network throughput. The algorithm reduced travel time by about 17–48% compared to conventional control.

Lee et al. (*12*) proposed an algorithm called "Cumulative Travel Time Responsive" control. The algorithm considers the accumulated travel time of vehicles entering each approach to the intersection (to a maximum distance of 300 m), and selects the phase with the highest combined travel time for service. The authors found that the method reduced total delay by 34%, with 100% CV penetration; they found that about 30% penetration was needed for improvement.

Xu et al. (*13*) proposed a cooperative method to optimize signal timing and trajectories of connected autonomous vehicles simultaneously. This study considered the sensing range. Extending the communication beyond 800 m did not result in improved performance, but the performance was found to degrade substantially when the range was reduced below 400 m.

One commonality in these algorithms is that they use a planning-horizon approach that seeks to find the optimal sequence and timing of phases over a timeline on the order of minutes. Such an approach is generally needed to manage the computational needs of the optimization processes.

**Self-Organizing Control**

Another class of control algorithm considers only the control of local intersections, with the rule-sets for that control anticipated to achieve emergent coordination. The concept of self-organizing signal control was explored by Gershenson et al (*14*, *15*), which demonstrated a remarkable level of coordination for some rather idealized networks, using a set of five rules for switching between two phases. One of the most important rules considers the number of vehicles approaching the intersection. Day and Bullock (*16*) explored the performance of this algorithm under constraints including the use of a range equivalent to 5 seconds of travel time for counting approaching vehicles. Delay was reduced up to about 38–56% under certain conditions.

Cesme and Furth (*17*) proposed the use of a "secondary extension" to implement self-organizing control into existing actuated traffic signal control. That is, a decision can be made to extend the green of a phase based on a table of arrivals for vehicles up to 20 seconds upstream from the intersection. Their method achieved dynamic coordination for a variety of networks, reducing delay in both under saturated and oversaturated conditions compared to conventional control.

An advantage of such methods is that the computational overhead is reduced, since the algorithms consider the control of only the local intersection. One may speculate that some previously documented single-intersection algorithms, perhaps going back as far as Miller's 1963 work (*18*), might possess such potential. Such methods could be integrated with data from AIS or CVs through construction of the arrival table with a feed of measured vehicle positions, rather than estimates based on point detection.

**METHODOLOGY**
This study identifies the impact of the sensing range on the performance of real-time control algorithms. Two algorithms with very different concepts of operations are adopted for this concept. First, the self-organizing algorithm (SOA) proposed by Cesme and Furth (*17*) is implemented. This method acts as an additional layer of intelligent actuation on a framework of otherwise conventional fully-actuated control. We also explore the performance of the Phase Allocation Algorithm (PAA) (*7*), which seeks to identify the optimal control decision over a planning horizon. Both algorithms utilize arrival tables that are developed from vehicle positions within some defined sensing range, which was adjusted in the simulation to establish the relationship between algorithm performance and sensing range.

**Self-Organizing Algorithm (SOA)**
The method uses conventional actuated control, into which a secondary extension has been added to provide coordination by allowing vehicles to extend the green under certain conditions.

*Efficient Actuated Control*
This study uses a fixed phase sequence, minimum green, maximum green, non-simultaneous gap out, and stop bar detectors to improve actuated signal control performances. Stop bar detectors 40 ft in length were placed on every phase, and advanced detectors were located 330 ft away from the stop bar for through movements along the arterial. Passage time for stop bar and advance detectors are set to be 2.1 and 3.0 seconds, respectively. Lane-by-lane detection was set up, and a flashing yellow arrow configuration for the left turn movements was assumed. A minimum green time of 5 seconds was used on each phase. A maximum green time of 35 seconds was used on the minor phases, with 60 seconds used on the arterial through phases.

*Secondary Extension Rules*
The SOA tries to extend the green time of the major through movements when a platoon is arriving at the intersection. This strategy tends to keep platoons together, but the tradeoff is that that delay increases on the minor phases, which must stay red for longer while the through movement remains green. To avoid causing excessive delays for non-arterial phases, some constraints are imposed on secondary extension.

The set of arriving vehicles determines the allowable secondary extension time, favoring more densely packed arrivals. The allowable lost time per vehicle, $L_v^*$, is the ratio of wasted green time during secondary extension to the number of vehicles arriving during the extension. This ratio is minimized by considering different potential lengths of secondary extension.

Let time *t* be initialized so that $t = 0$ at the gap out time, and let $L_v^*(t)$ be the lost time per vehicle if the secondary extension's length is *t*, given by

$$L_v^*(t) = \frac{(t - n(t) h_{\text{sat}})}{n(t)} \qquad (1)$$

where, *t* is the sensing range in seconds, *n(t)* is the number of vehicles expected to pass the stop line if the green phase is further extended by *t*, $h_{sat}$ and is the saturation headway. $L_v^*(t)$ is calculated for discreet values of t up to $SX_{max}$, the maximum allowed length of secondary extension, to find an optimal value of $t$, thus:

$$L_v^* = \min_{t \leq SX_{max}} L_v(t) \qquad (2)$$

A secondary extension is less likely to occur as the overall level of intersection utilization increases. The affordable lost time, $L_a$, is defined by

$$L_a = \min\left[2, 2\frac{1}{X_c - 1}\right] \qquad (3)$$

The calculation of $X_c$ is described in detail in (*19*). Our tests used lost time = 4 s per critical phase, and measured $X_c$ adaptively (updating estimates for every 5 cycles). The maximum allowable value for $SX_{max}$ is 30 s. Let $\Delta c_{nj}$ represent the difference in effective cycle length with a neighboring intersection $j$. The maximum allowable green extension time is given by

$$SX_{max} = \min\{\max(10, \Delta C_{n1}, \Delta C_{n2}), 30\} \qquad (4)$$

Secondary extension is allowed only once per cycle per arterial phase.

**Phase Allocation Algorithm (PAA)**
The PAA runs a bi-level optimization to minimize total vehicle delay on the basis of estimated vehicle arrivals. At the upper level, a dynamic program uses a forward and backward recursion. The forward function passes the barrier lengths to the lower level, which estimates the total vehicle delay as the summation of queue length of each phase over time. The forward function also records the optimal values for each stage. The backward function retrieves the optimal signal plan starting from the last stage. Barrier sequences are fixed but phase sequences are flexible. The algorithm optimizes the timing plan only once at the beginning of each barrier group and projects the timing plan for the next cycle.

To describe the algorithm, we must define several terms: a set of phases, $P$; the total number of discrete time steps, $T$; the minimum green time, $g$; the summation of the amber and red time, $r$; the control variable $x_j$, minimum barrier length, $x_j^{min}$; maximum barrier length, $x_j^{max}$; denoting the length of stage $j$; the set of feasible control decisions $X_j(s_j)$, given state $s_j$; the travel delay $f_j(s_j, x_j)$; and the cumulative value of prior travel delay, $v_j(s_j)$.

The process of forward recursion is summarized in Algorithm 1. In the DP, for each stage calculation, forward function calculates the allocated time to each barrier group. The algorithm continues to plan for as many barriers as possible until two consecutive barrier group experiences same total delay for each time step (line 7 in Algorithm 1).

Given state $s_j$, the feasible range for control variable is

$$x_j(s_j) = \begin{cases} 0 & s_j < x_j^{min} \text{ or } T-s_{j-1} \leq x_j^{min} \\ x_j^{min} \cdots x_j^{max} & s_j \geq x_j^{min} \text{ and } T - s_{j-1} - s_1 \geq x_j^{max} \\ x_j^{min} \cdots T - s_{j-1} - s_1 & s_j \geq x_j^{min} \text{ and } x_j^{min} \leq T - s_{j-1} - s_1 \leq x_j^{max} \end{cases} \qquad (5)$$

**ALGORITHM 1 Forward Recursion**

1:    $j \leftarrow 1$ and $v_j(0) \leftarrow 0$
2:    **for** $s_j = 1, ...., T$ **do**
3:        $v_j(s_j) = \min \{ f_j(s_j, x_j) + v_{j-1}(s_{j-1}) \mid x_j \in X_j(s_j) \}$
4:        Record $x_j^*(s_j)$ as the optimal solution
5:    **end for**
6:
7:    **if** $j \geq 3$ and $v_{j-1}(T) == v_j(T)$ **then** STOP
8:    **else** $j = j+1$

To calculate the optimal performance measure at forward recursion a lower level utility optimization is necessary. The lower level function minimizes vehicle delay for different combinations of phase sequences and phase durations based on the given barrier length and barrier state. Vehicle delay is defined as summation of queue length. Queue length at time step *n,* depends on queue length at previous time step *n* – 1, arrival and departure flow at time step *n*. Details of the lower level value function calculation may be found in the original study (*7*).

After the forward recursion, optimum results for each stage will be retrieved through backtracking. The retrieval policy is summarized in Algorithm 2.

**ALGORITHM 2 Backward Recursion**

1:    $s_{j-1}^* \leftarrow T$
2:    **for** $j = J-1, J-2, ....1$ **do**
3:        $s_{j-1}^* = s_j^* - x_j^*(s_j)$
4:    **end for**

The algorithm considers coordination as a fixed priority request and includes the request as an added constraint within the dynamic programming framework. Phase split time for only arterial phases are considered as fixed priority requests.

**Implementation**

The control logic was implemented in a Python script using the Component Object Model (COM) interface with the VISSIM simulation software. Two different Python scripts were coded to execute SOA and PAA. These scripts took sensing range, VISSIM network, and signal configurations as input and executed the VISSIM file externally. In every time step, VISSIM sends each vehicle's position and speed data to the script. Vehicles that are outside of the sensing range have been filtered out from the list for purposes of modeling the sensing range limitations. The link and lane positions of the vehicles are used to identify the corresponding intersection and movement type.

Virtual Econolite ASC/3 controllers were used to control the signals. For SOA, secondary extensions were implemented by placing a hold on the target phase. Otherwise, the intersections operated under ordinary fully-actuated control. For implementation of PAA, the intersections were completely divorced from the detector inputs, and the timing of phases was determined solely by the timing decisions supplied by PAA, implemented through placement of phase calls in the controller.

**Simulation Scenarios**

A hypothetical corridor with three intersections was modeled in VISSIM, with distances between intersections shown in Figure 1. The intersections were operated as eight-phase intersections with protected-permitted left turns. Each street had four lanes, and every intersection included left-turn lanes on all four approaches. A speed of 45 mph was used.

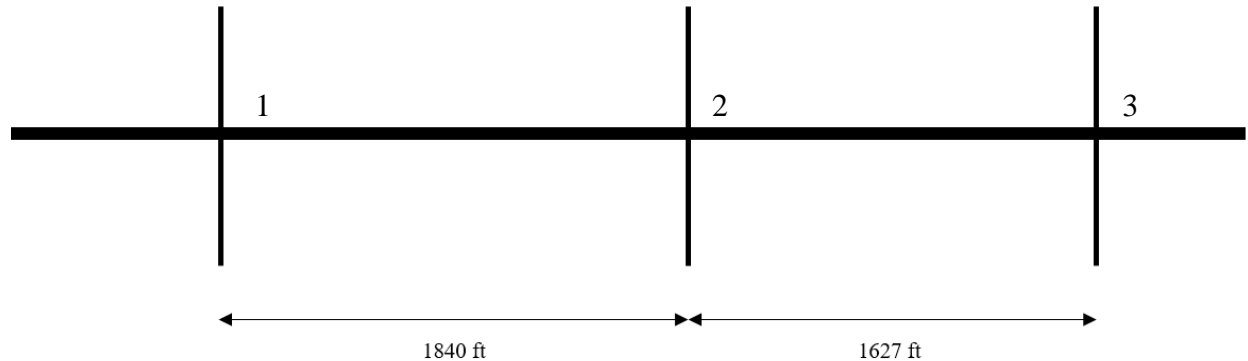

**FIGURE 1 Simulation network used in this study**

Three volume scenarios were developed, as follows:

• *Scenario 1, Symmetric volume*. Eastbound and westbound movements have similar volumes, and cross-street traffic is substantially lower than the main street. Each intersection's degree of utilization is approximately 75%.
• *Scenario 2, Asymmetric volume*. The eastbound through volume is significantly higher than the westbound through volume. Similar to symmetric volume scenario, cross-street traffic in this scenario is substantially lower than the arterial. Intersection degree of utilization values are 70–80%.
• *Scenario 3, Balanced volume*. All the through movements (northbound, southbound, eastbound and westbound) have similar volumes. Intersection degree of utilization values are 70–80%.

The scenarios are compared with conventional fully-actuated and coordinated-actuated control. The VISTRO software program was used to determine the cycle, offsets, and splits for a coordinated-actuated control pattern. In VISTRO, the signal timing was optimized using the Hill Climbing algorithm using the Performance Index using a factor of 0.2 to convert the number of stops into the equivalent amount of total delay in hours. The optimization process considered cycle lengths between 60 and 120 seconds.

For sensitivity analysis, six different sensing ranges of 165, 330, 660, 1000, 1500 and 2000 ft were tested. Five simulation runs were carried out for each of these scenarios for SOA and PAA, as well as for the two baseline control scenarios.

**RESULTS AND ANALYSIS**

**Total Delay**

Figure 2 shows the total system delay for all scenarios, algorithms, and sensing ranges considered. Each chart shows the relationship between performance and the sensing range, as compared to two conventional control methods: coordinated-actuated and fully-actuated non-coordinated control, which are represented by flat lines since those methods of control did not use any AIS enhancements for which the sensing range is relevant.

Both algorithms have the highest delay at the smallest sensing range, with delay gradually decreasing as the sensing range is increased. SOA tended to have lower total delay than PAA except at the longest sensing ranges. Both PAA and SOA yielded lower total delay than coordinated-actuated control even at shorter sensing ranges (with the exception of the smallest ranges for PAA in some cases), while they did not surpass fully-actuated control except in the longest sensing ranges. In terms of total delay, both SOA and PAA appear to be able to outperform fully-actuated control within a sensing range between 1000 and 1500 ft, with the exception that under symmetric volumes the total delay under SOA is similar to that of fully-actuated control and does not surpass it.

There is no change in the performance of SOA within 330 ft because within that distance the green is extended by the "primary" extension from the advance detector located at 330 ft. The secondary extension is only called if the primary extension is inactive. In a few cases, slow moving vehicles (mostly vehicles making right turns) can induce a secondary extension, but the effects are negligible. At the longest sensing ranges considered here, the total delay under SOA sometimes starts to increase. This is because at the longest sensing ranges, longer maximum allowable secondary extensions have to be used, causing non-arterial movements to wait longer, thereby increasing the total delay of all movements.

The performance of PAA improves with longer sensing ranges. In the symmetric scenario, its performance converges at 1500 ft, whereas for the asymmetric and balanced scenarios there is a very slight decrease in delay when the sensing range is further extended to 2000 ft. Under the lowest sensing ranges, PAA sometimes performs worse than coordinated-actuated control. At 165 ft, the sensing range is equivalent to 3 seconds of travel time, and only information from vehicles arriving within the next 3 seconds is available, which is clearly not adequate for the algorithm to make control decisions. In summary, PAA achieves the lowest delay of the two algorithms but requires a longer sensing range to accomplish this, while SOA has the lowest delay under shorter sensing ranges.

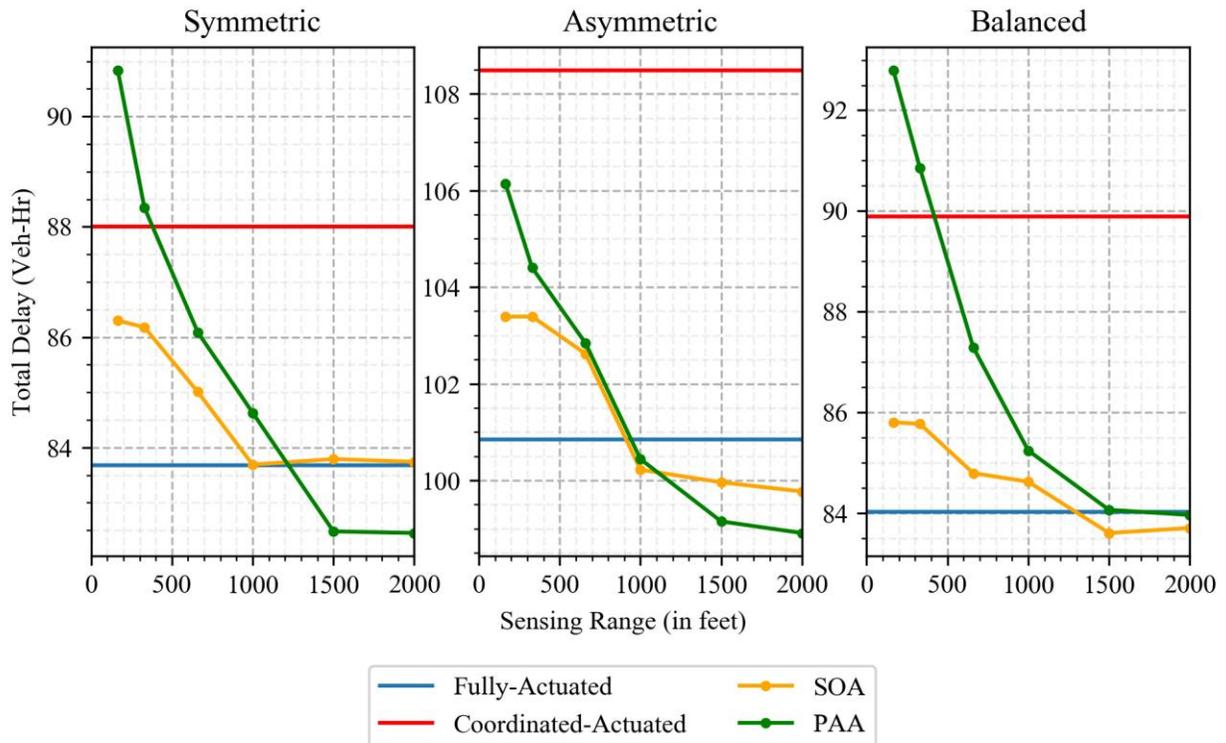

**FIGURE 2 Total delay results by volume scenario**

**Arterial and Non-arterial Delay**

The previous analysis used the performance of fully-actuated and coordinated-actuated control to bracket the performance of the algorithms in terms of total delay, and it was observed that both SOA and PAA were only able to achieve lower total delay than fully-actuated control under the longest sensing ranges. This is not surprising because fully-actuated control is a very efficient mode of control for a single intersection in terms of total delay. However, in signal control the operational objective is usually not only to achieve the lowest total delay, but instead some movements are prioritized. Increases in non-arterial movement delay are often accepted to achieve lower delay for arterial movements, even though in most cases this leads to higher total delay.

To explore this further, Figure 3 plots the performance of SOA, PAA, fully-actuated control, and coordinated-actuated control in terms of delay for the non-arterial movements (vertical axis) and the arterial movements (horizontal axis). For the SOA and PAA data, the darker color represents the longer sensing ranges.

The difference in the positions of fully-actuated and coordinated-actuated control illustrate the tradeoff inherent in the implementation of coordination. Coordination yields a lower amount of arterial delay, at the cost of higher non-arterial delay, as would be expected since the arterial movements are prioritized over the non-arterial movements. The performance of SOA and PAA falls somewhere in between fully-actuated and actuated-coordinated control.

SOA is closer to fully-actuated control, which is not surprising since the method consists of secondary extensions that are layered on top of fully-actuated control. SOA moves the performance closer to coordinated-actuated control, achieving lower arterial delay at the expense of increases in non-arterial delay. With respect to the sensing range, the results tend to form a V-shaped pattern. That is, as the sensing range increases there is an improvement in both arterial and non-arterial delay to a certain point. After this point, if the sensing range is increased further, the arterial delay is reduced but the non-arterial delay increases.

PAA offers similar operation to coordinated-actuated control, but usually has lower non-arterial delay, and with longer sensing ranges it can also yield lower arterial delay. Depending on the completeness of the estimated arrival table, PAA can provide lower arterial delay since it can sometimes provide longer green times than the phase splits without significant impact on non-arterial movements. Additionally, among the four methods considered here, only PAA uses flexible phase sequencing. Even with these control features, SOA and fully-actuated control produced lower non-arterial delay, although at the longest sensing ranges PAA can sometimes approach the performance of PAA.

These results suggest that signal control algorithms of either type likely fall somewhere between the two "poles" of standard performance represented by fully-actuated and actuated-coordinated control. Naturally, it would be ideal to develop algorithms that can move the performance into the lower left-hand quadrant. However, because coordination under any mechanism forces some vehicles to wait while others take priority, it is not possible to obtain something for nothing.

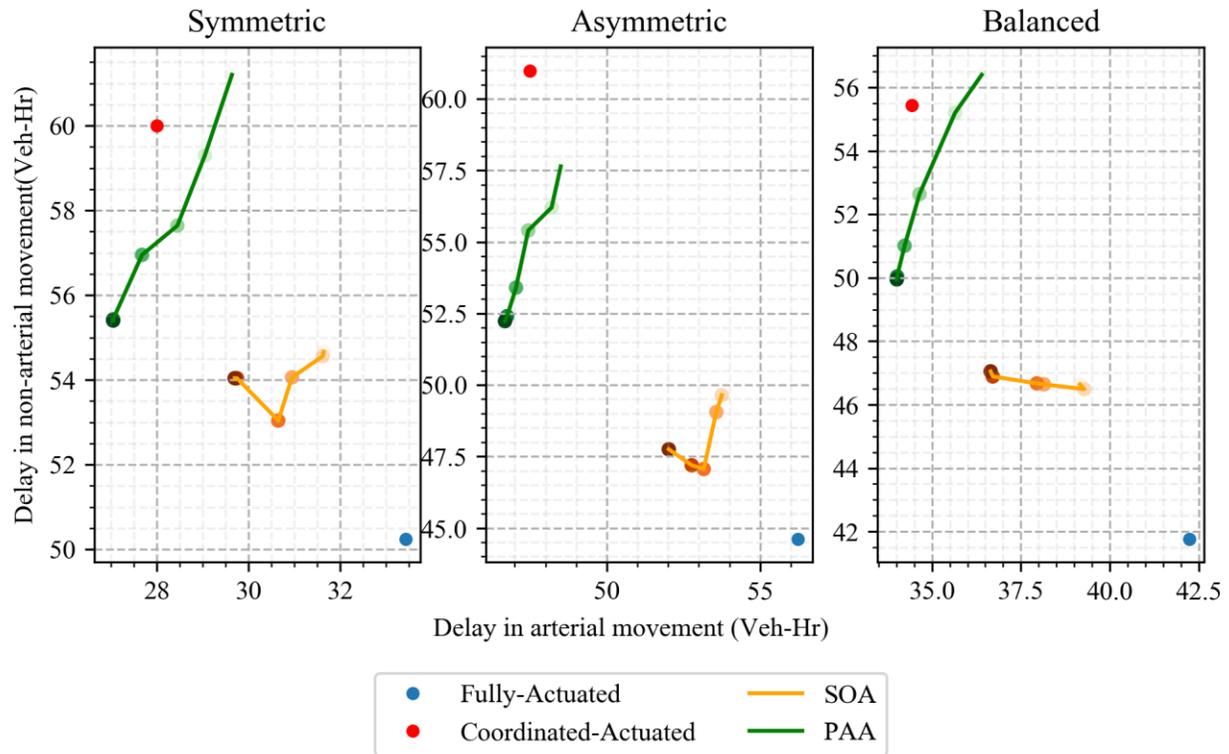

**FIGURE 3 Arterial vs. non-arterial delay comparison**

**Travel Times**

Figure 4 shows cumulative frequency distributions of travel times for eastbound and westbound through traffic for the symmetric volume scenario.

Across all scenarios, coordinated-actuated control and PAA provided lower median travel times than SOA and fully-actuated control. In addition, the slopes of the curves are steeper for coordinated-actuated control and PAA, indicating more reliable travel times. This is not surprising, since the latter two provide more aggressive mechanisms for coordination. In some cases PAA yields superior performance to coordinated-actuated control. At lower sensing ranges, PAA performs similar to fully-actuated control.

SOA had lower median travel times than actuated-control in most but not all cases. In asymmetric-eastbound and balance-westbound corridor SOA actually yields the longest travel time. This is likely attributable to "lapses" in coordination where the arterial green is unable to start at an appropriate time because non-arterial traffic is taking longer to finish service, which itself may be a consequence of implementing secondary extension. It seems that the self-organizing potential was not fully realized under that particular scenario. Under other scenarios, SOA is able to marginally improve the travel time.

Regarding the impact of sensing range, PAA exhibits decreases in travel time as the sensing range is increased, for the most part. The longest sensing ranges yield the lowest median travel times. A similar trend is observed for SOA, but it is less consistent.

**Quality of Progression**

Figure 5 depicts Coordination Diagrams (*20*) for the eastbound through movement at intersection 2. These diagrams show the patterns of vehicle arrivals relative to the phase state at the intersection, with each "column" representing an effective signal cycle. The figure depicts operating conditions under the symmetric volume scenario for various sensing ranges under PAA and SOA. The blue bars in the SOA diagrams indicate when the secondary extension is implemented. At the top of the figure, the results for fully-actuated control and coordinated-actuated control are shown.

Under SOA, the maximum allowable secondary extension is constrained by the arriving platoon. The arriving platoon is further constrained by the sensing range. At lower sensing ranges, almost no platoons receive any benefit from secondary extension, but as the sensing range is increased, the maximum allowable secondary extension increases and the extensions are implemented more frequently.

Under PAA, the effective cycle length and duration of green are similar to coordinated-actuated control, as illustrated by similar appearance of the red and green intervals in the coordination diagrams. It also appears that PAA occasionally terminates the green for the through movement early, as shown by some cycles where the intervals are much shorter than others. This occurs when there are few vehicles approaching the intersection at the time when such a decision is made. For PAA with a sensing range of 165 ft, this situation was common since only few vehicles are observable within such a short range. Other than this occasional short-cycle behavior, PAA exhibits rather consistent operation from one cycle to the next, which is not surprising given that the coordinated phases are served as a fixed priority request. Though the PAA is not explicitly maintaining a cycle length, priority requests from coordinated phases are cyclic in nature. Hence cycle length in PAA is not a constant value like coordinated-actuated control but it maintains significant level of consistency.

Figure 5 also depicts the percentage of green (POG) on arrival for the corresponding coordination diagram. From the POG, coordinated-actuated control and PAA shows better performance in terms of coordination than actuated and SOA. Though, both SOA and PAA shows better POG with longer sensing ranges, the relationship is not exactly linear and smooth. POG values suggest that though the aggregate results show improvement, individual intersection movement may experience inconsistency.

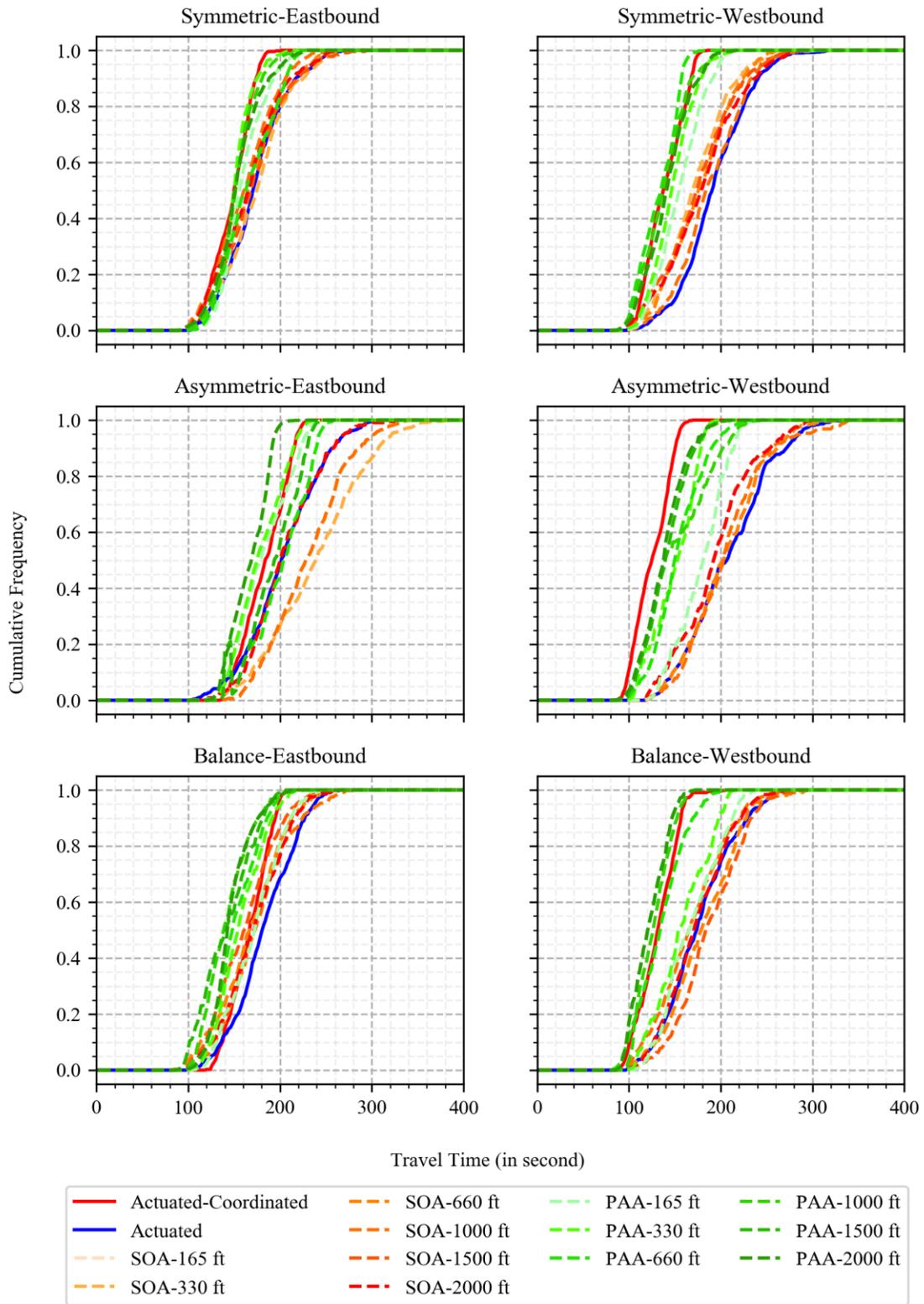

**FIGURE 4 Cumulative frequency diagram for corridor travel times for all scenarios**

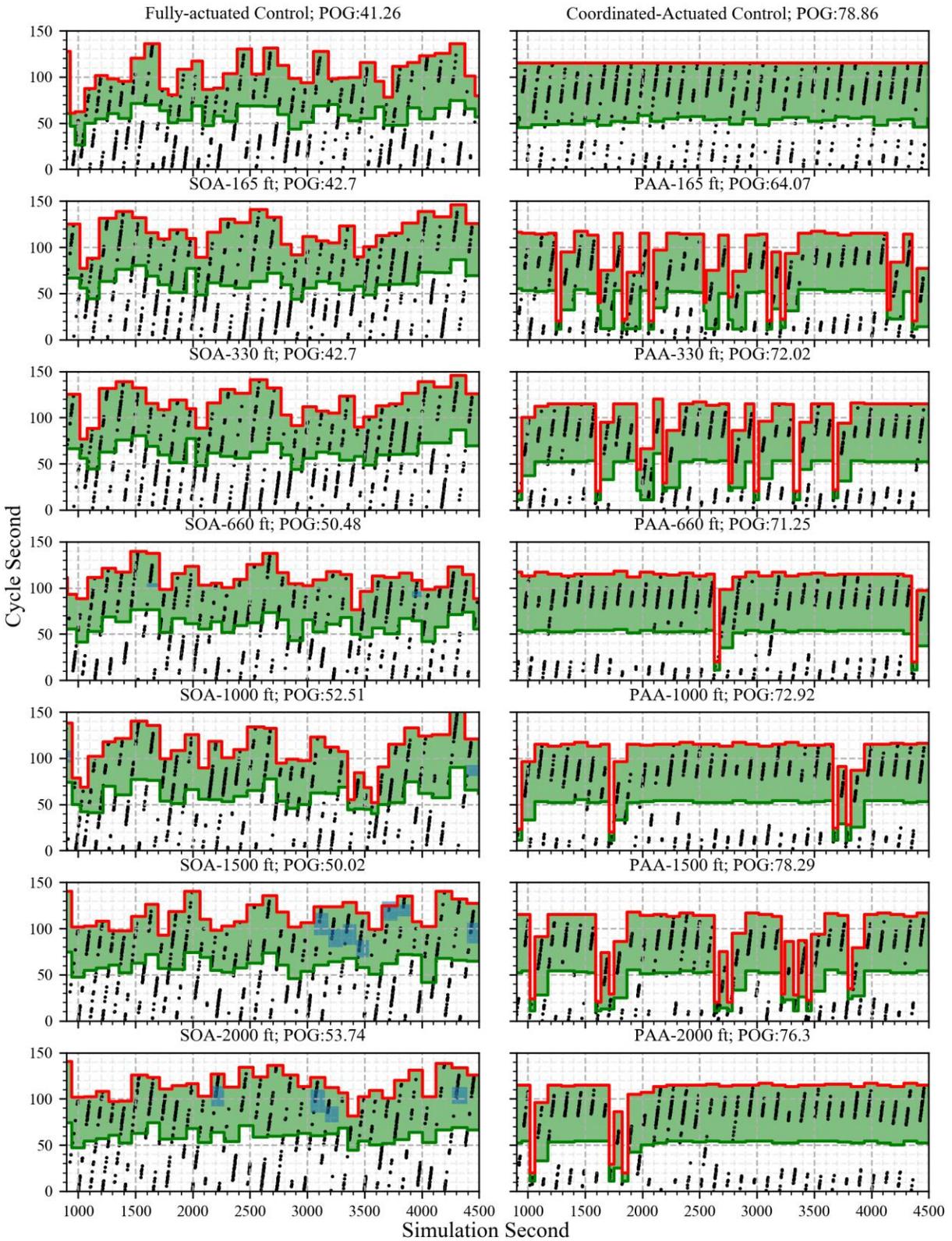

**FIGURE 5 Coordination diagram comparison for intersection 2 and east bound through traffic**

**IMPLICATIONS FOR SENSING RANGE NEEDS**

It is expected that a trajectory-based signal control algorithm should be able to improve its performance with a longer sensing range, up to a certain point where that performance converges, and additional distance provides no further benefit. Improvements in performance under shorter sensing ranges do not necessarily guarantee that the performance will improve with larger sensing ranges. In some cases, as seen in the results here, adding sensing range can lead to increases in delay. Inconsistent performance suggests that the algorithm is not able to use additional trajectory data effectively, or the decision process may favor the arterial movement to the detriment of the total intersection delay.

For the lowest sensing ranges, neither algorithm performed well. SOA under sensing range of 330 ft or lower merely replicated the results of fully-actuated signal control with an advance detector, and there was no added benefit from implementing secondary extension. These shorter sensing ranges were also insufficient for PAA to make better control decisions, although PAA outperformed coordinated-actuated control in the asymmetric scenario.

Both SOA and PAA demonstrated improved performance as the sensing range increased. If the sensing range was between 330 ft to 1000 ft, both algorithms improved and outperformed coordinated-actuated control. Although fully-actuated control and SOA experienced similar total delays near 1000 ft, SOA provides slightly better travel time characteristics along the major arterial.

For sensing ranges greater than 1000 ft, SOA sometimes showed inconsistent performance which can be attributed to the tradeoff between arterial and non-arterial movement delay. For the balanced volume scenario, SOA outperformed the other methods. The performance of PAA converged after 1500 ft and showed lowest total delay in symmetric and asymmetric scenario. For sensing ranges of 1000 ft or more, SOA and PAA both appear to provide tangible improvements over conventional methods, but which of the two is more appropriate further depends on the volume configuration as well as the operational objectives. Sensing ranges of about 1000 ft were found to be needed to match or surpass the amount of total delay under fully-actuated control.

**CONCLUSIONS**

This study evaluated the performance of two algorithms that use vehicles speed and position data under different spatial ranges. Outcomes were examined for a variety of traffic scenarios representing various conditions for a three-intersection corridor with two-phase intersections. A self-organizing algorithm (SOA) was compared along with the Phase Allocation Algorithm (PAA) against fully-actuated and coordinated-actuated control.

This study focused on the impact of the sensing range. In terms of total delay, both SOA and PAA yielded better results as the sensing range was increased, which is the expected result since a long sensing range means a greater number of arriving vehicles are observed. However, the results tended to converge at a particular distance, meaning that no further improvement could be obtained by extending the range beyond that distance. For the symmetrical volume scenario, the SOA results converge at 1000 ft and PAA converges at 1500 ft. In the asymmetric scenario, neither algorithm converged within 2000 ft, but the decrease in delay beyond 1500 ft was very marginal. In the balanced scenario, the performance of PAA was again only marginally improved beyond 1500 ft, while the performance of SOA actually got a little worse from going from 1500 to 2000 ft.

Comparing the delays of arterial and non-arterial movements casts further light on the nature of the algorithms. As the sensing range is increased, PAA reduces delays for both types of movements, until its performance converges. Meanwhile, SOA reduces delays for both types of movements up to a certain point, beyond which the delay for the arterial movements is decreased while the non-arterial delay is increased. If more aggressive coordination is desired, PAA may be a desirable algorithm to employ, as it is able to achieve similar arterial delay to coordinated-actuated control while providing added flexibility to decrease non-arterial delay. On the other hand if the objective is to achieve lower total delay while providing a looser mechanism to provide coordination as the opportunity arises, then SOA would appear to be appropriate.

The travel time performance provides another point of comparison between SOA and PAA. The PAA is able to achieve lower and more reliable travel times, surpassing actuated-coordinated control in many cases. SOA yields travel times that are closer to fully-actuated control, better in some cases and worse in others. The results tend to favor one direction over another.

Coordination diagram and Percentage of Green (POG) on arrival data for symmetric scenario suggests that, both the algorithm will benefit from longer sensing range, however the relation between sensing range and percentage of green is not always consistent even though the aggregate results for all intersections like travel time showed more consistency.

These results suggest that advanced algorithms such as SOA and PAA may be capable of achieving better performance than conventional methods, particularly with longer sensing ranges. The results of this study would suggest that about 1000-1500 ft would be needed to provide performance surpassing actuated control in terms of total delay. Under shorter distances in between 660-1000 ft, both SOA and PAA are capable of providing improvements over coordinated-actuated control. SOA with 1000 ft of sensing range provides similar delay compared to fully-actuated control, but can provide better and more reliable travel times along the arterial. For the shortest considered sensing ranges (between 165-330 ft), SOA performed similar to fully-actuated control while PAA sometimes had greater total delay than coordinated-actuated control.

**ACKNOWLEDGMENT**

This work was sponsored by the National Renewable Energy Laboratory, operated by Alliance for Sustainable Energy, LLC, for the U.S. Department of Energy (DOE) under Contract No. DE-AC36-08GO28308. Funding was provided by the DOE Vehicle Technologies Office (VTO) under the Systems and Modeling for Accelerated Research in Transportation (SMART) Mobility Laboratory Consortium, an initiative of the Energy Efficient Mobility Systems (EEMS) Program. The views expressed in the article do not necessarily represent the views of the DOE or the U.S. Government. The U.S. Government retains and the publisher, by accepting the article for publication, acknowledges that the U.S. Government retains a nonexclusive, paid-up, irrevocable, worldwide license to publish or reproduce the published form of this work, or allow others to do so, for U.S. Government purposes.

**REFERENCES**
1. Day, C. M., R. Helt, D. Sines, and A. M. T. Emtenan. Leveraging Sensor-Based Vehicle Position and Speed Information in Traffic Signal Control with Existing Infrastructure. *2019 IEEE Intelligent Transportation Systems Conference, ITSC 2019*, 2019, pp. 4049–4054. https://doi.org/10.1109/ITSC.2019.8917351.
2. Datesh, J., W. T. Scherer, and B. L. Smith. Using K-Means Clustering to Improve Traffic Signal Efficacy in an IntelliDriveSM Environment. *2011 IEEE Forum on Integrated and Sustainable Transportation Systems, FISTS 2011*, 2011, pp. 122–127. https://doi.org/10.1109/FISTS.2011.5973659.
3. He, Q., K. L. Head, and J. Ding. PAMSCOD: Platoon-Based Arterial Multi-Modal Signal Control with Online Data. *Procedia - Social and Behavioral Sciences*, Vol. 17, No. December, 2011, pp. 462–489. https://doi.org/10.1016/j.sbspro.2011.04.527.
4. Sen, S., and K. L. Head. Controlled Optimization of Phases at an Intersection. *Transportation Science*, Vol. 31, No. 1, 1997, pp. 5–17. https://doi.org/10.1287/trsc.31.1.5.
5. Mirchandani, P., and L. Head. A Real-Time Traffic Signal Control System: Architecture, Algorithms, and Analysis. *Transportation Research Part C: Emerging Technologies*, Vol. 9, No. 6, 2001, pp. 415–432. https://doi.org/10.1016/S0968-090X(00)00047-4.
6. Goodall, N., B. Smith, and B. Park. Traffic Signal Control with Connected Vehicles. *Transportation Research Record*, No. 2381, 2013, pp. 65–72. https://doi.org/10.3141/2381-08.
7. Feng, Y., K. L. Head, S. Khoshmagham, and M. Zamanipour. A Real-Time Adaptive Signal Control in a Connected Vehicle Environment. *Transportation Research Part C: Emerging Technologies*, Vol. 55, 2015, pp. 460–473. https://doi.org/10.1016/j.trc.2015.01.007.
8. Feng, Y., M. Zamanipour, K. L. Head, and S. Khoshmagham. Connected Vehicle-Based Adaptive Signal Control and Applications. *Transportation Research Record*, Vol. 2558, 2016, pp. 11–19. https://doi.org/10.3141/2558-02.

9. Beak, B., K. Larry Head, and Y. Feng. Adaptive Coordination Based on Connected Vehicle Technology. *Transportation Research Record*, Vol. 2619, 2017, pp. 1–12. https://doi.org/10.3141/2619-01.
10. Wang, P. (Slade), P. (Taylor) Li, F. R. Chowdhury, L. Zhang, and X. Zhou. A Mixed Integer Programming Formulation and Scalable Solution Algorithms for Traffic Control Coordination across Multiple Intersections Based on Vehicle Space-Time Trajectories. *Transportation Research Part B: Methodological*, Vol. 134, 2020, pp. 266–304. https://doi.org/10.1016/j.trb.2020.01.006.
11. Islam, S. M. A. B. Al, and A. Hajbabaie. Distributed Coordinated Signal Timing Optimization in Connected Transportation Networks. *Transportation Research Part C: Emerging Technologies*, Vol. 80, 2017, pp. 272–285. https://doi.org/10.1016/j.trc.2017.04.017.
12. Lee, J., and B. Park. Development and Evaluation of a Cooperative Vehicle Intersection Control Algorithm under the Connected Vehicles Environment. *IEEE Transactions on Intelligent Transportation Systems*, Vol. 13, No. 1, 2012, pp. 81–90. https://doi.org/10.1109/TITS.2011.2178836.
13. Xu, B., X. J. Ban, Y. Bian, W. Li, J. Wang, S. E. Li, and K. Li. Cooperative Method of Traffic Signal Optimization and Speed Control of Connected Vehicles at Isolated Intersections. *IEEE Transactions on Intelligent Transportation Systems*, Vol. 20, No. 4, 2019, pp. 1390–1403. https://doi.org/10.1109/TITS.2018.2849029.
14. Gershenson, C. Self-Organizing Traffic Lights. *Complex Systems*, Vol. 16, 2005, pp. 29–53.
15. Gershenson, C., and D. Rosenblueth. Self-Organizing Traffic Lights at Multiple-Street Intersections. *Complexity*, Vol. 17, No. 4, 2011, pp. 23–39. https://doi.org/10.1002/cplx.
16. Day, C. M., and D. M. Bullock. Investigation of Self-Organizing Traffic Signal Control with Graphical Signal Performance Measures. *Transportation Research Record*, Vol. 2620, No. 2620, 2017, pp. 69–82. https://doi.org/10.3141/2620-07.
17. Cesme, B., and P. G. Furth. Self-Organizing Traffic Signals Using Secondary Extension and Dynamic Coordination. *Transportation Research Part C: Emerging Technologies*, Vol. 48, 2014, pp. 1–15. https://doi.org/10.1016/j.trc.2014.08.006.
18. Alan, M. J. A Computer Control System for Traffic Networks. In *Proceedings of the 2nd International Symposium on the Theory of Traffic Flow* (J. Almond, ed.), Elsevier, New York, 1963, pp. 200–220
19. Day, C. M., D. M. Bullock, and J. R. Sturdevant. Cycle-Length Performance Measures: Revisiting and Extending Fundamentals. *Transportation Research Record*, No. 2128, 2009, pp. 48–57. https://doi.org/10.3141/2128-05.
20. Day, C. M., R. Haseman, H. Premachandra, T. M. Brennan, J. S. Wasson, J. R. Sturdevant, and D. M. Bullock. Evaluation of Arterial Signal Coordination: Methodologies for Visualizing High-Resolution Event Data and Measuring Travel Time. *Transportation Research Record*, No. 2192, 2010, pp. 37–49. https://doi.org/10.3141/2192-04.